\documentclass[conference, 10pt, a4paper]{IEEEtran}
\IEEEoverridecommandlockouts
\usepackage{cite}
\usepackage{amsmath,amssymb,amsfonts}
\usepackage{subcaption,caption}
\usepackage{algorithmic}
\usepackage{graphicx}
\usepackage{textcomp}
\usepackage{xcolor}
\usepackage{soul}
\usepackage{braket}

\def\BibTeX{{\rm B\kern-.05em{\sc i\kern-.025em b}\kern-.08em
    T\kern-.1667em\lower.7ex\hbox{E}\kern-.125emX}}

\begin{document}

\title{A Survey and Tutorial on Security and Resilience \\of Quantum Computing\\
\thanks{The work is supported by 
National Science Foundation (NSF) (CNS- 1722557, CCF-1718474, OIA-2040667, DGE-1723687 and DGE-1821766) and seed grants from Penn State Institute for Computational and Data Sciences (ICDS) and Huck Institute of the Life Sciences.

We acknowledge the use of IBM Quantum services for this work. The views expressed are those of the authors, and do not reflect the official policy or position of IBM or the IBM Quantum team.}
}

\makeatletter
\newcommand{\linebreakand}{%
  \end{@IEEEauthorhalign}
  \hfill\mbox{}\par
  \mbox{}\hfill\begin{@IEEEauthorhalign}
}
\makeatother

\author{
\IEEEauthorblockN{Abdullah Ash Saki}
\IEEEauthorblockA{\textit{School of EECS} \\
\textit{Pennsylvania State University}\\
University Park, PA \\
axs1251@psu.edu}
\and
\IEEEauthorblockN{Mahabubul Alam}
\IEEEauthorblockA{\textit{School of EECS} \\
\textit{Pennsylvania State University}\\
University Park, PA \\
mxa890@psu.edu}
\and
\IEEEauthorblockN{Koustubh Phalak}
\IEEEauthorblockA{\textit{School of EECS} \\
\textit{Pennsylvania State University}\\
University Park, PA \\
krp5448@psu.edu}
\linebreakand
\IEEEauthorblockN{Aakarshitha Suresh}
\IEEEauthorblockA{\textit{School of EECS} \\
\textit{Pennsylvania State University}\\
University Park, PA \\
ams9647@psu.edu}
\and
\IEEEauthorblockN{Rasit Onur Topaloglu}
\IEEEauthorblockA{
\textit{IBM Corporation}\\
Hopewell Junction, NY \\
rasit@us.ibm.com}
\and
\IEEEauthorblockN{Swaroop Ghosh}
\IEEEauthorblockA{\textit{School of EECS} \\
\textit{Pennsylvania State University}\\
University Park, PA \\
szg212@psu.edu}
}

\maketitle
\thispagestyle{plain}
\pagestyle{plain}

\begin{abstract}
Present-day quantum computers suffer from various noises or errors such as, gate error, relaxation, dephasing, readout error, and crosstalk. Besides, they offer a limited number of qubits with restrictive connectivity. Therefore, quantum programs running these computers face resilience issues and low output fidelities. The noise in the cloud-based access of quantum computers also introduce new modes of security and privacy issues. Furthermore, quantum computers face several threat models from insider and outsider adversaries including input tampering, program misallocation, fault injection, Reverse Engineering (RE) and Cloning. This paper provides an overview of various assets embedded in quantum computers and programs, vulnerabilities and attack models and the relation between resilience and security. We also cover countermeasures against the reliability and security issues and present future outlook for security of quantum computing.
\end{abstract}



\maketitle

\section{Introduction}
\label{sec:introduction}
Quantum computing is a paradigm-shifting technology that can revolutionize fields like drug discovery, chemistry~\cite{kandala2017hardware}, machine learning~\cite{qml}, and optimization. Quantum computing is evolving rapidly from a theoretical concept to the experimental demonstration of quantum advantage~\cite{arute2019quantum}. Various vendors like IBM~\cite{ibmqx}, Amazon~\cite{aws-braket}, and Microsoft~\cite{azure} are offering cloud-based access to physical quantum computers. Several physical hardware are being developed by IBM, Pasqal, Rigetti, Xandau and Google. Software packages such as, Qiskit, Circ, Orquestra, and Forest are also available to design, optimize, and execute quantum algorithms. With progress on hardware and software, quantum computing is poised to meet the scalability requirements to solve practical size problems. 

However, present-day quantum computers suffer from several challenges which include a small number of qubits, restrictive connectivity, and limited native instructions. Besides, the qubits are error-prone and have short lifetimes. Quantum circuits are affected by these errors and generate low-quality outputs. Therefore, noise-resilient methods are of paramount importance. These devices are often termed as noisy intermediate-scale quantum (NISQ) computers~\cite{preskill}. Apart from engineering better qubits, several other independent approaches like parameter optimization in variational algorithms, noise-aware mapping, and scheduling are pursued to make quantum programs resilient to noise. 
\begin{figure*}
    \centering
    \includegraphics[width=0.80\textwidth]{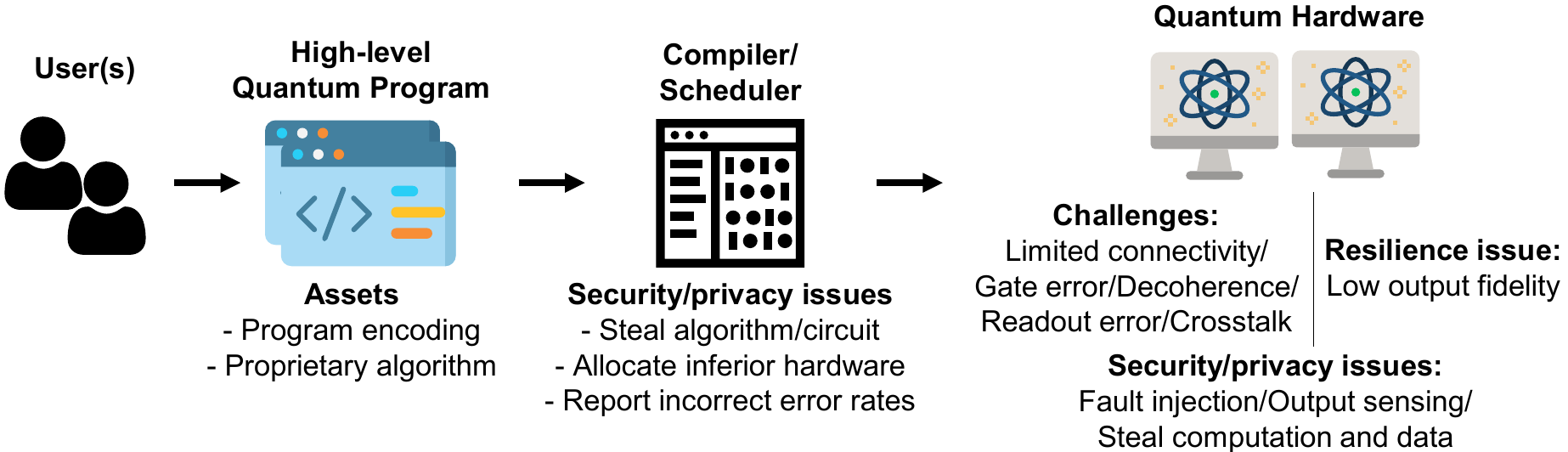}
    \caption{Overview of assets, challenges and security/privacy issues in quantum computing. (Attribution: quantum hardware icon made by Wichai.wi from www.flaticon.com and other icons made by Freepik from www.flaticon.com)}
    \label{fig:overview}
\end{figure*}
For example, parameterized quantum circuits (PQCs) are proposed for variational algorithms and machine learning~\cite{farhi2014quantum, dallaire2018quantum, kandala2017hardware, schuld2020circuit}. 
Resilience of a variational algorithm, namely, quantum approximate optimization algorithm (QAOA) under noise, and compilation techniques to achieve better resilience have been studied in \cite{alam-cicc, alam-iccad-invited-qaoa, alam-micro, mahabubul-dac-qaoa, alam-date}. The resiliency of various quantum factoring algorithms have been investigated in ~\cite{ling-vqf}. A hybrid quantum-classical approach is proposed ~\cite{saki-trng} to counteract noise in NISQ based true random number generator. The above approaches achieve noise resiliency through parameter optimization. Besides parameter optimization, various software techniques like mapping and scheduling are being explored that aim to minimize the number of SWAPs for communication~\cite{zulehner-a*}, allocation of gates to less noisy qubits~\cite{tannu, qure, murali-noise-adaptive}, reducing crosstalk through intelligent scheduling~\cite{murali-crosstalk}, or blend outputs of different mappings to sanitize errors~\cite{tannu2019ensemble}. In general, the methods try to improve the performance of quantum algorithms by mitigating numerous errors.

In addition to resilience issues, NISQ devices and the cloud model of quantum computing can be susceptible to many security and privacy challenges. For example, quantum circuits can be treated as intellectual properties (IP). An untrusted software chain or server can steal data and computation and cause undue loss to users. Besides, hardware architectures with 1000 qubits are projected in less than 5-years. Such scale makes multi-tenant computing~\cite{multi-programming} lucrative in quantum clouds as it can increase device utilization and commercial gains. However, multi-tenant access to quantum hardware can lead to fault injection~\cite{saki-islped} and data-leakage attacks~\cite{saki-tqe-sensing}. Moreover, malicious entities at the cloud-end can allocate inferior hardware and/or report incorrect error rates. Running on inferior devices or qubits can lead to low program fidelity. 

In this paper, we (i) review various challenges and security vulnerabilities in NISQ computers; (ii) review various resilience techniques proposed in the literature to mitigate errors; and (iii) discuss attack models, countermeasures, and security opportunities relevant to NISQ architectures.

In the remaining of the paper: Section~\ref{sec:challenges} presents the challenges and security vulnerabilities present in devices. Section ~\ref{sec:attack-models} describes attack models demonstrated in NISQ devices. Section~\ref{sec:resilience} reviews various error mitigation methods for noise resilient performance. Section~\ref{sec:countermeasures} presents several countermeasures to thwart attacks. Section~\ref{sec:opportunities} describes security opportunity in terms of true random number generator (TRNG). Section~\ref{sec:outlook} presents the future outlook on resilience and security of quantum computers. Finally, Section~\ref{sec:conclusion} draws conclusion. 

\section{Challenges and vulnerabilities}\label{sec:challenges}
\subsection{Qubit lifetime and errors}
Quantum bits or qubits on NISQ devices suffer from a short lifetime and erroneous gate and measurement operations. A short qubit lifetime entails a spontaneous loss of qubit state (saved data) which is called \emph{decoherence}. For example, a qubit in state $\ket{1}$ interacts with the environment, spontaneously loses energy, and ends up with state $\ket{0}$. This phenomenon is known as \emph{relaxation}. Another qubit lifetime issue is \emph{dephasing} where a qubit state loses its phase information. Relaxation and dephasing are quantified by T1 and T2 times, respectively.
Besides, the gate and measurement operations on qubits are erroneous. These are known as \emph{gate error} and \emph{measurement/readout error}, respectively. The logical output of a quantum gate may be incorrect due to the readout error. Similarly, measurement error can flip a qubit state (i.e., may be recorded as 0 instead of 1 and vice versa). Gate error and measurement error are quantified by gate error rate and measurement error rate, respectively.

Moreover, crosstalk errors can be present in the device due to which two parallel gate operations can experience increased gate error. In~\cite{sarovar2019detecting}, the authors presented multiple sources of classical crosstalk that can be present in Transmon-based quantum computers such as, traditional electromagnetic (EM) crosstalk between
microwave lines, stray on-chip EM fields, etc. On a high level, the signal intended to control one qubit can disturb another independent qubit. For example, in~\cite{flux-line-crosstalk} the authors demonstrate one such case where control fields (magnetic flux) used to operate qubits affect unaddressed qubits. The work in~\cite{saki-tqe} reported that including crosstalk in quantum circuit simulation could lead to more accurate results. It establishes crosstalk as a prevailing source of error in NISQ devices.
Finally, all the errors (decoherence, gate error, measurement error, and crosstalk) show temporal variation. 

\subsection{Limited native gates}
A quantum program or circuit can be constructed with any high-level quantum gate (any reversible gate can be a quantum gate). However, the NISQ devices support only a few native gates. For example, there are 4 single-qubit gates (`id', `rz', `sx', and `x') and only 1 two-qubit gate (`cx') in IBM machines. Any high-level gate is decomposed to the native gates of the hardware. The decomposition step increases the number of gate counts and run-time of the circuit. Increased gate count and run-time adversely affect circuit performance due to short qubit lifetime and gate error.

\subsection{Coupling constraint}\label{subsec:coupling-constraint}
\begin{figure}
    \centering
    \includegraphics[width=2.8in]{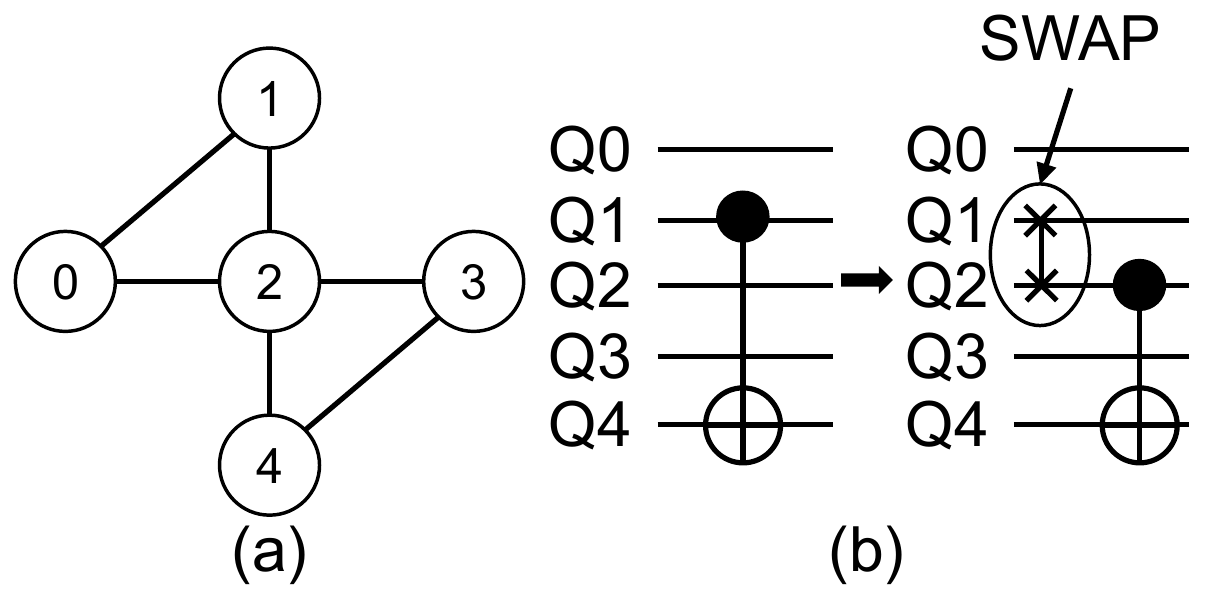}
    \caption{(a) The coupling map of ibmq\_5\_yorktown. (b) A SWAP gate is inserted to satisfy coupling constraint as Q1 and Q4 are not connected.}
    \label{fig:coupling-constraint}
\end{figure}
NISQ devices, especially quantum computers based on superconducting qubits, have limited connectivity between qubits known as \emph{coupling constraint}. The limited connectivity prevents 2 qubit gates between any two arbitrary qubits. A compiler needs to add additional SWAP operations to satisfy the coupling constraint. Fig.~\ref{fig:coupling-constraint}a shows the coupling graph of an IBM quantum computer where we cannot perform a CNOT (CX) gate directly between Q1 and Q4 as they are not connected. One option is to \emph{SWAP} Q1 and Q2 so that the data of Q1 moves to Q2. Now, the CX can be applied between Q2 and Q4 as they are connected. A SWAP operation includes 3 CX gates (when translated to the native instructions of the device) that has two implications. First, it increases the run-time of the quantum program. Second, it increases gate counts. Increased run-time and gate counts negatively affect the circuit performance due to the aforementioned qubit lifetime and gate error issues. The SWAP inserted circuit that satisfies the coupling constraint is termed as a nearest-neighbor (NN) compliant circuit.

\subsection{Cloud-based access}\label{subsec:cloud}
Quantum computing is far from becoming a personal commodity due to cost and needs for ultra-cold temperature~\cite{krantz2019quantum, bruzewicz2019trapped}, shielded environment, and complex wiring for control. Therefore, cloud-based access to quantum computers is the logical path forward where the hardware is hosted in a remote location.
In addition to hardware, high-performance quantum simulators also need cloud offerings as personal computers can hit memory limit while simulating a reasonably large circuit. 


Quantum computing in the cloud is growing at a tremendous rate in the past few years. IBM is providing free access to its quantum processors and simulators through the cloud (IBM Quantum Experience). Rigetti, another quantum computing vendor, started a similar service named Quantum Cloud Service or QCS (which is now retired and moved to AWS Braket and Azure Quantum). Companies such as, Microsoft, Google, Amazon, D-Wave, Xanadu, IonQ, etc. are also providing various services for quantum computing in the cloud.

With cloud offering being the only option, various security and privacy issues can emerge. For example, a malicious entity on the cloud-end can assign an inferior hardware~\cite{puf} or report incorrect error-rates~\cite{samah-iccad} (discussed in Section~\ref{sec:attack-models} as ``scheduler attacks''). Besides, a rogue element in the cloud can steal the structure and/or the output of a quantum program which can be intellectual properties (IPs). For example, a company can invest substantial time and money to design a quantum algorithm to solve a difficult problem (e.g., drug discovery). These make it lucrative for an adversary to observe the submitted quantum programs and steal information.

\subsection{Problem encoding}
The construction of a quantum circuit can reveal sensitive information about the problem. We discuss one such vulnerability using a combinatorial problem MaxCut which involves dividing a graph into two parts so that the maximum number of edges is cut. Such problems can be formulated using the spin-glass/Ising model. Fig.~\ref{fig:problem-encoding} (right) shows the quantum approximate optimization algorithm (QAOA) circuit for the problem graph in Fig.~\ref{fig:problem-encoding} (left). Here, each edge in the graph is represented by a CNOT-rotation-CNOT gate in the circuit. Therefore, one can infer the problem graph just by looking at the circuit. Besides MaxCut, many Quadratic Unconstrained Binary Optimization (QUBO) problems fall under this category.

\begin{figure}
    \centering
    \includegraphics[width=3.2in]{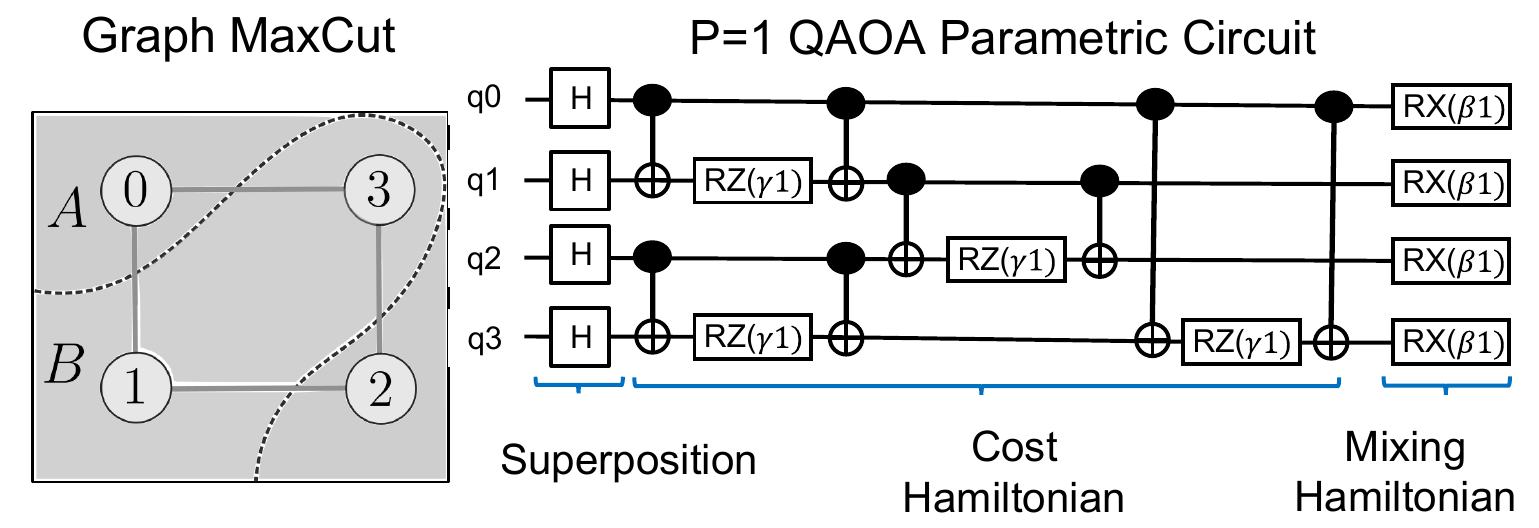}
    \caption{(left) An example graph for MaxCut (right) QAOA circuit for finding MaxCut of the graph on the left.}
    \label{fig:problem-encoding}
\end{figure}


\subsection{Need for untrusted compilers}
One of the important aspects of quantum circuit compilation is to optimize the circuit for improved circuit depth and reduced gate count. Several $3^{rd}$ party compilers are evolving that offer optimization at faster compilation time even for large quantum circuits \cite{zulehner-a*, muqut}. The following factors will motivate the quantum circuit designers to avail the untrusted $3^{rd}$ party compilation services: (a) success of quantum circuit since the optimized circuit is essential to obtain meaningful results from NISQ computers. A poorly optimized circuit will produce random outputs even if it is functionally identical; (b) lack of trusted compilers that have caught up with the latest advancements in optimization; (c) availability of efficient but untrusted $3^{rd}$ party compilers\cite{multi-programming, zulehner-a*, muqut} that are being developed to optimize depth and gate count compared to trusted compilers.
These compilers can be hosted on either the local machines by the $3^{rd}$ party or on the cloud service providers to launch, (i) cloning/counterfeiting, where the quantum circuit can be stolen or reproduced; and (ii) Reverse Engineering (RE), where the sensitive aspects of the quantum circuit could be extracted.

\subsection{Application oriented vulnerabilities: quantum machine learning}
Quantum machine learning (QML) is an emerging field that aims to develop quantum algorithms to perform conventional generative/discriminative machine learning tasks (e.g., classification, regression, etc.) \cite{qml, killoran2019continuous, dallaire2018quantum, schuld2020circuit}. 
An extremely high-dimensional \emph{classical} data can be loaded into a few qubits using methods like amplitude encoding~\cite{mottonen2004transformation}.
For example, $2^n$ classical features can be encoded as the amplitudes of different basis states in an $n$-qubit system. Therefore, QML has been explored with to provide an advantage over the existing classical ML algorithms for high-dimensional workloads (e.g., object detection in high-resolution images without dimension reduction). 

Recent research on QML has uncovered several performance bottlenecks. %
The work in \cite{mcclean2018barren} revealed that the training landscape in parameterized quantum circuits might have vanishing gradients. These locations of vanishing gradients are referred to as $barren~plateaus$ in the literature. Once stuck in a $barren~plateau$, gradient-based optimization methods (e.g., stochastic gradient descent) may not be able to move further to train the network. In \cite{wang2020noise}, the authors showed that quantum-noise could also induce $barren~plateaus$ in the PQC training landscape. In \cite{mahabubul-islped}, the authors demonstrated that the temporal variation in quantum-noise can affect the reliability of a quantum classifier. In \cite{liu2020vulnerability}, the authors demonstrated that a small amount of perturbation in the input data is enough to induce misclassification in a trained quantum classifier. As the required perturbation scales inversely with the dimension, a small perturbation is sufficient to induce misclassification in high-dimensional quantum classifiers. An adversary can exploit these performance bottlenecks or vulnerabilities to attack QML applications. Several attacks have been demonstrated already in recent academic studies \cite{saki-islped, lu2020quantum}. 


\section{Attack models}\label{sec:attack-models}
\subsection{Security attacks}
\subsubsection{Crosstalk induced fault injection}
Crosstalk error can be exploited to launch fault-injection attacks in a multi-tenant computing (MTC) environment where two or more quantum programs may run simultaneously on a different set of physical qubits. MTC is economically enticing as it maximizes hardware resource usage and profit.
\begin{figure}
    \centering
    \includegraphics[width=2.7in]{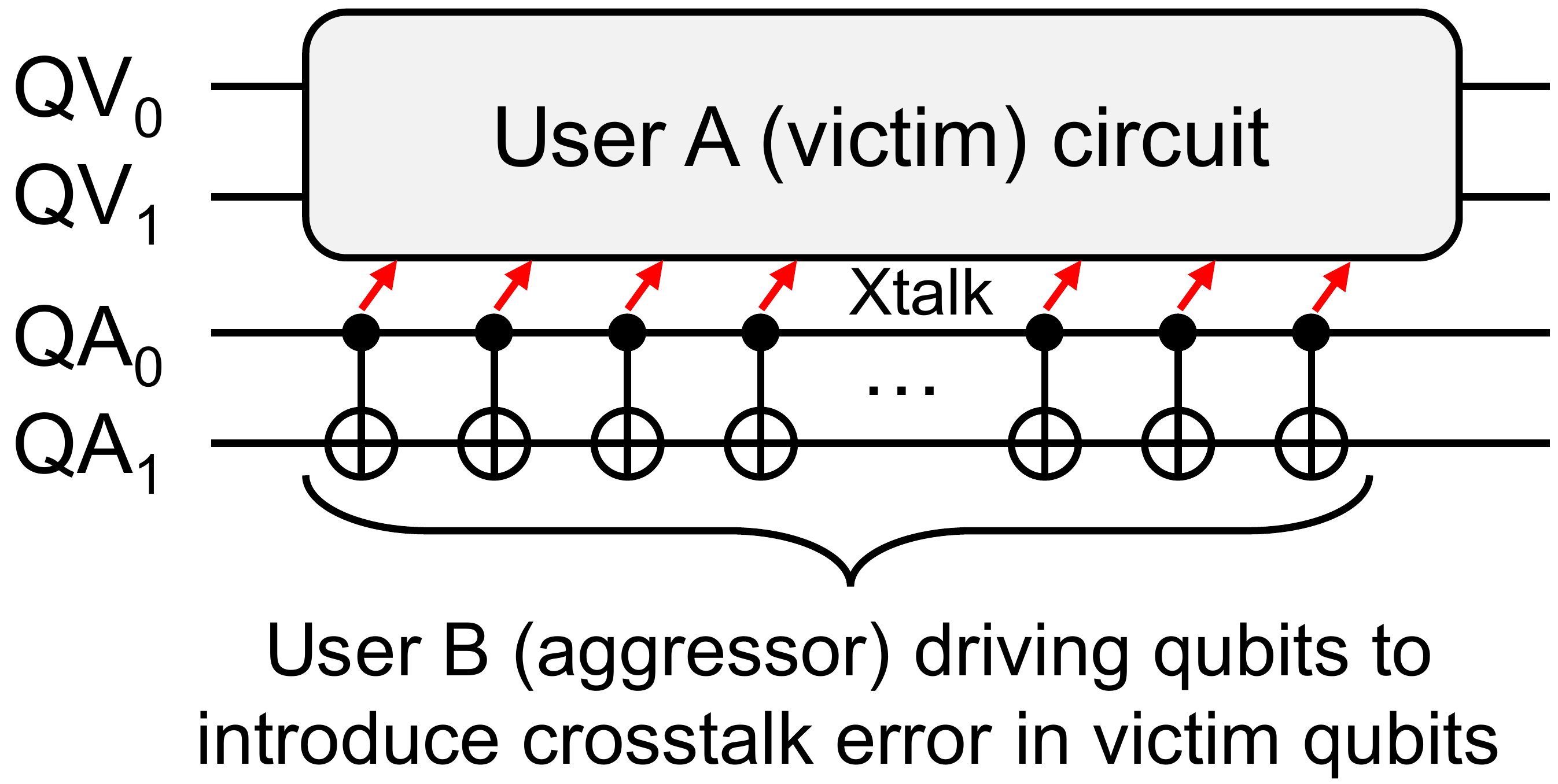}
    \caption{A conceptual diagram of crosstalk induced fault-injection attack.}
    \label{fig:fault-injection}
\end{figure}
Fault-injection in the computation process may have substantial socio-economical significance. For example, a deterministic fault in the weather forecast or optimal power grid topology calculation can provide undue financial/political advantage to the adversary. The attack model from~\cite{saki-islped} assumes that the adversary can run his/her program in the same hardware as one or more victim programs. A conceptual diagram is given in Fig.~\ref{fig:fault-injection}. They assume that the adversary, (i) will know the public information e.g., coupling map of the hardware; (ii) may also be aware of the crosstalk values between various qubits by running crosstalk characterization experiments such as, idle tomography and simultaneous randomized benchmarking on the qubits before the attack. 
\begin{figure}
    \centering
    \includegraphics[width=3in]{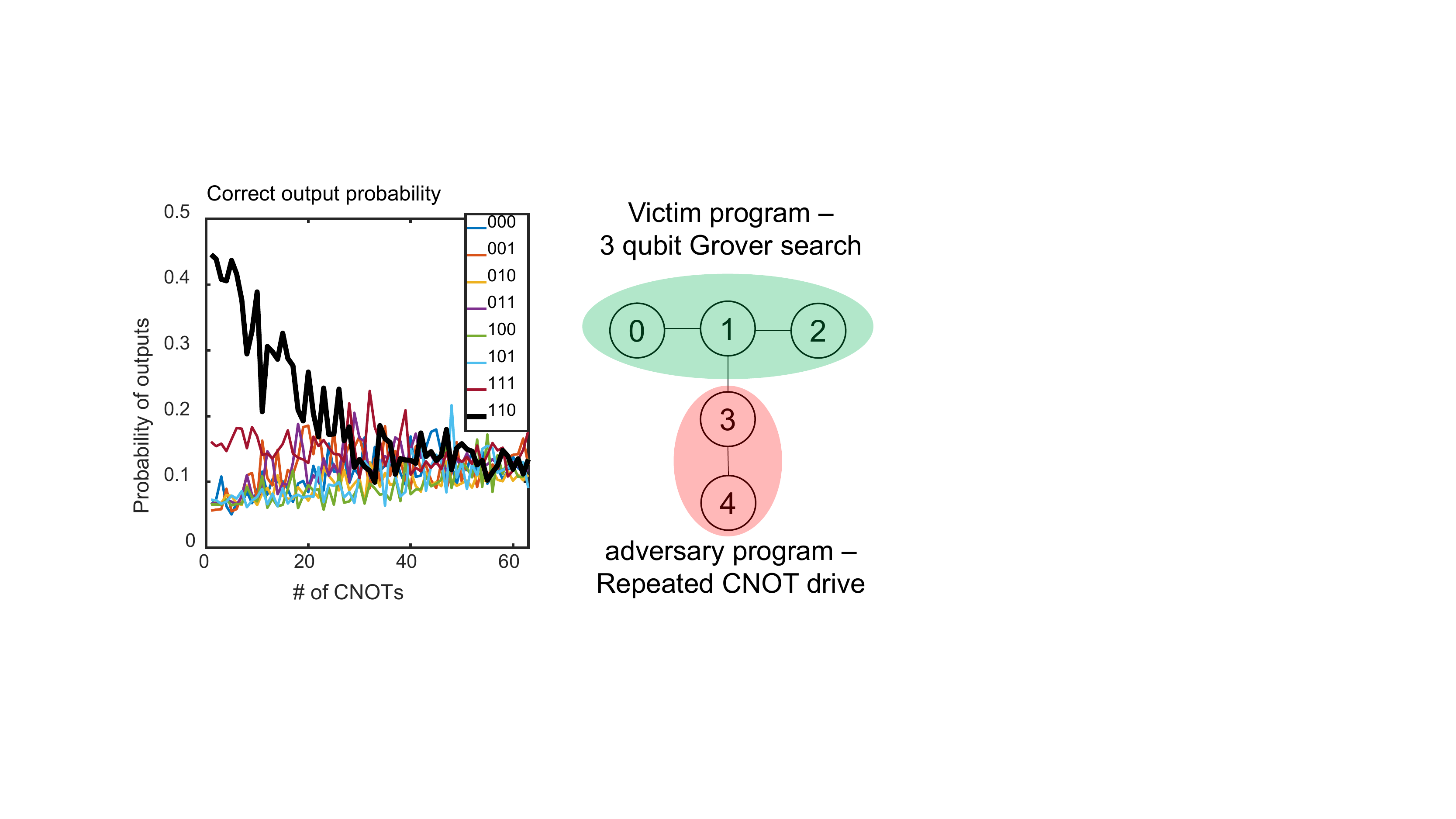}
    \caption{3 qubit Grover search under attack.}
    \label{fig:grvr-3}
\end{figure}
The authors demonstrated the attack using both simulation and experiments. They picked a 3-qubit Grover search as the victim program and repeated CNOT drive as the adversary program. The results, collected from \emph{ibmq\_5\_yorktown (ibmqx2)} a Canary processor, (Fig.~\ref{fig:grvr-3}) show that the correct output probability of the Grover search falls drastically (i.e., becomes indistinguishable) after a certain number of CNOTs in the adversary circuit.  
\subsubsection{Scheduler attacks}
Quantum circuits are sent to quantum hardware via a cloud-based provider which allocates the hardware for the circuit. Here, the user has no visibility on the hardware that is being allocated. In \cite{puf}, the authors propose a new attack model where the user is allocated an inferior quantum hardware instead of the desired one. 
Moreover, even if the desired quantum hardware is allocated, the scheduling policy for the queue of quantum circuits is another aspect that should be taken into consideration. 
The queue of quantum programs on the cloud side is usually long, with the main goal of maximizing throughput for cost reduction and better scientific exploration. The scheduling policy of the hardware is usually provided by the vendor for program allocation to hardware. In Fig. \ref{fig:attack_model}, two users \textbf{U1} and \textbf{U2} request for hardware \textbf{A} which is better compared to hardware \textbf{B} in terms of characteristics like error rates and fidelity. First U1 sends the request to the cloud service, and cloud service allocates hardware A to U1. However, when U2 requests for hardware B, the cloud service could make U2 either wait or allocate hardware B, which is of inferior quality. If the latter happens, U2 will suffer from incorrect results due to inferior hardware and may also end up paying more.

A variant of the scheduler attack is discussed in~\cite{samah-iccad} where the scheduler allocates the requested device however, a rogue employee in the quantum computing company can be the attacker. He/she can alter the reported error-rate data so that an inferior qubit (with a larger error rate) is reported as a superior qubit (with a smaller error rate). When a user requests or a compiler allocates physical qubits, they (user or compiler) can unknowingly select inferior quality qubits to run the circuit. Therefore, the user circuit will experience heightened errors and sub-optimal output. The authors state that the success of the attack depends on the variation of the error rates of the underlying hardware and the failure model applied to determine the mapping policy. An attack can be triggered if there is a significant variation in the qubit error rates. 

\subsubsection{Attacks on QML}

Similar to the attack models on classical ML algorithms, attacks on QML can be categorized from three dimensions- (i) timing i.e., if the attack take place during training or inference; (ii) information i.e., the type of information that is available to the attacker e.g., knowledge of the internals of the QML model/algorithm (white-box attacks) or access to the inputs/outputs of the QML model only (black-box attacks); and (iii) goals i.e., the objectives of the attacker e.g., force misclassification for certain inputs (targeted attacks) or affect the overall reliability of a model (non-targeted attacks). In \cite{saki-islped}, the authors demonstrated a non-targeted/reliability attack where an attacker induces noise to a victim's quantum classifier (during inference) through crosstalk in a MTC environment. It increased the misclassification rate of the classifier significantly. 
In \cite{lu2020quantum}, the authors demonstrated ways to generate adversarial samples for a QML image classifier (noisy inputs that are misclassified by the classifier) in both white-box and black-box setup. The additional noise acted as a unitary that modifies the input state to the classifier. However, using such adversarial samples to perform actual attacks is an open research question.

\begin{figure}[htbp]
\centering
 \includegraphics[width=0.8\linewidth]{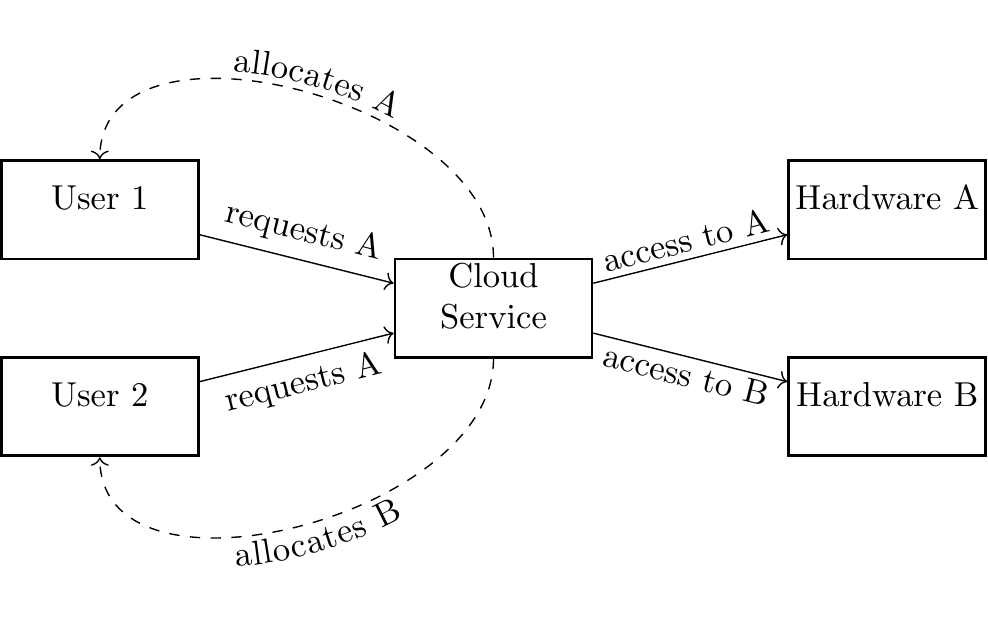}
 \caption{Scheduler attack model~\cite{puf}.}
 \vspace{-0.2em}
 \label{fig:attack_model}
\end{figure}

\subsection{Privacy attacks}
\subsubsection{Readout sensing}
The readout error in a quantum computer is state-dependent which means state $\ket{1}$ and $\ket{0}$ experience asymmetric bit-flip probabilities. Besides, the asymmetry extends beyond a single qubit. For example, if 2 qubits are read, then 4 possible states - $\ket{00}$, $\ket{01}$, $\ket{10}$, and $\ket{11}$ - will show asymmetric bit-flip probabilities. Therefore, if the two qubits belong to two different programs each, one adversary and another victim, the adversary can \textit{sense} the state of the victim just by reading his/her qubit. The work in \cite{saki-tqe-sensing} demonstrated that an adversary can exploit readout error and infer another user's output by reading his/her qubit. 

The authors provide experimental results that show the adversary output distribution is glaringly different for victim qubit being $\ket{0}$ and $\ket{1}$. The sensing attack involves two steps, (i) the adversary collects reference signatures from a device by running circuits on both qubits, (ii) the adversary reads only his/her qubit and compares it with the reference signature using a statistical distance (Jensen-Shannon Distance). If the collected signature is statistically closer to reference signature $\ket{1}$ than $\ket{0}$ then it is inferred as $\ket{1}$ and vice-versa. The authors report an inferencing accuracy of $96\%$ from experiments. 

\subsection{Vulnerabilities in reversible circuits}
Quantum Circuits are based on reversible logic. There are several works on the security of reversible circuits~\cite{samah-iccad-ic-ip, reversible-trojan, ets-limaye, samah-ic-ip-piracy-tvlsi}. 
In~\cite{samah-ic-ip-piracy-tvlsi}, an IP/IC piracy attack for reversible circuits was presented along with countermeasures. A target Boolean logic function can be converted into a reversible circuit using synthesis such as, quantum multiple-valued decision Diagrams (QMDD), binary decision diagram (BDD), etc. The synthesis step adds ancillary lines on the input side and garbage lines on the output side of the logical function.
Reversible circuits have a certain level of inherent privacy as an adversary needs to know which are ancillary and garbage lines and values of ancillary bit to identify the functionality.
However, the synthesis approach can leave ``telltale'' signs that help an adversary to locate ancillary and garbage lines. The adversary can exploit these left-over signs to launch an attack which entails knowing the embedded functionality of the circuit.
In~\cite{ets-limaye}, the authors consider the \emph{end-user} to be untrusted along with the foundry. They show that an end-user can launch piracy and reverse engineering attacks with access to a netlist and a functional chip.


\section{Resilience Techniques}\label{sec:resilience}
\subsection{Depth optimal qubit mapping and routing}
As mentioned in Section~\ref{subsec:coupling-constraint}, additional SWAP gates, that are necessary to satisfy coupling constraint, are detrimental for circuit performance. Thus, it is desired to keep the SWAP gates to a minimum. As the problem is NP-complete~\cite{siraichi2018qubit}, researchers often resort to heuristic techniques to make it scalable. 

In~\cite{zulehner-a*}, the authors exploited A* search to find a depth-optimal version of the NN-compliant circuit. They defined a cost function $f(x) = g(x) + h(x) + l(x)$ where $g(x)$ is the path cost, $h(x)$ is the heuristic cost, and $l(x)$ is the look-ahead cost. The algorithm takes one layer of the circuit. The physical qubits used in the circuit and their neighbors are considered for SWAP insertion. For each possible SWAP, a heuristic cost $h(x)$ is computed. The cost is the new distance between qubits after adding the SWAP (adding a SWAP usually minimizes the distance). Besides, adding a SWAP affects future layers. The lookahead cost $l(x)$ accounts for this. The SWAP with the lowest $f(x)$ is selected as the candidate SWAP, and the search continues. 

The proposal in~\cite{gushuli-tackling} argues that not all candidate qubits (used physical qubits in the circuit and their neighbors) have the same priority in the SWAP insertion decision. They only consider qubits used in the particular layer and their neighbors. This reduced set of qubits curtails the number of SWAPs to be checked, decreasing the search time in large quantum circuits significantly. 
Many other works exist~\cite{siraichi2018qubit, siraichi2019qubit, murali2019formal, temporal-planner, muqut, debjyoti, 45shafaei2013optimization, wille2019mapping, niu} on mapping and routing based on heuristic and exact methods. 

\subsection{Noise-aware mapping}
Gate error rates vary among qubits in real devices i.e., some qubits are better (less erroneous) to perform computations than others. This observation has led to several noise-aware mapping techniques~\cite{tannu, qure, murali-noise-adaptive}. The main idea is to prioritize less erroneous qubits to perform the majority of the gates. 
In~\cite{tannu}, the authors first proposed leveraging qubit-to-qubit variation to improve the program success rate. They proposed variation-aware qubit allocation (VQA) and variation-aware qubit movement (VQM) policies. In VQA, a set of physical qubits are picked to maximize their cumulative connectivity strength. The cumulative coupling strength is defined as the sum of success probabilities of all coupling links between the qubit and its neighbors. Its strength reflects two things: (i) a qubit is connected to more neighbors which is beneficial for optimal routing (less SWAP), and (b) the 2-qubit operations between the qubit and its neighbors will be less erroneous. Besides, the VQM policy ensures that the compiler will pick a routing path that uses fewer erroneous links. 

In~\cite{qure}, the authors started with a depth optimal NN-compliant version of the circuit using an algorithms as in~\cite{zulehner-a*} and searched for an isomorphic sub-graph from the device coupling graph with best program fidelity. The method contained the depth of the circuit (beneficial to counteract qubit lifetime issue) while finding better qubits and links to run the program.
In~\cite{murali-noise-adaptive}, the authors used satisfiability-modulo-theorem (SMT) to make qubit allocation and movement decisions while keeping error rate variations in mind. They also included readout error in their allocation decision besides gate error. Their weighted approach provided users with the flexibility to prioritize between gate and measurement errors. 

\subsection{Crosstalk mitigation}
The \emph{classical} version of the crosstalk arises when two gates run in parallel. Due to crosstalk, gate errors of two parallel gates can increase compared to isolated error rates. In~\cite{murali-crosstalk}, the authors conducted extensive experiments on several IBM devices to characterize crosstalk using simultaneous randomized benchmarking (SRB)~\cite{srb}. They had two key observations: (i) not all couplings are susceptible to crosstalk i.e., crosstalk between some couplings is negligible whereas some couplings can experience a 2X-3X increase in error rates due to crosstalk, and (ii) crosstalk becomes negligible after \emph{1-hop} distance. 

On the basis of crosstalk characterization results, the authors proposed a gate scheduling technique to minimize crosstalk. The core idea involved serializing the parallel gates at the cost of increased program depth (run-time) and hence, decoherence. The authors devised an SMT-based scheduler (\emph{XtalkSched}) to compare the reduction in crosstalk error and increase in decoherence error due to gate serialization. Then, the scheduler only serialized gates that provided an overall reduction in error from these two conflicting conditions. Up to 5.6X improvement in program fidelity has been reported due to crosstalk-aware scheduling.

Another work in~\cite{yding-xtalk} proposed dynamic assignment of qubit frequency to minimize crosstalk. Qubits are frequency addressable. If two neighboring qubits have sufficiently different operating frequencies, they will be less susceptible to crosstalk from another. However, the qubit operating frequency range is small which causes \emph{frequency crowding}. The authors presented a software solution to solve the frequency crowding issue and to dynamically allocate separate frequencies to neighboring qubits. They report 13.3X improvement in program success rate on average compared to gate serialization technique. The authors note that the solution is applicable to frequency tunable qubits.

\subsection{Readout error mitigation}
Once a quantum circuit completes execution, its final state is measured. In the superconducting quantum hardware from IBM, a single-qubit measurement is done by (i) running the readout pulse on the readout channel of the qubit, and (ii) recording the corresponding signal, which measures the energy state of the qubit, on the acquire channel. The signal recorded during the entire acquire duration is summed to generate a single complex value which is then plotted in an I-Q plane where $\ket{0}$ and $\ket{1}$ are supposed to show distinct clusters. Currently, IBM uses a linear classifier to classify the measured state ($\ket{0}$ or $\ket{1}$) from the imaginary and the real components of the complex value. The classifier is trained using a synthetic dataset. To generate this dataset, the qubit is prepared in the $\ket{0}$ and $\ket{1}$ state multiple times followed by measurement operations. The real and imaginary components associated with each measurement are used as the input features to the classifier. The actual states are used as the labels.

In \cite{patel2020disq}, the authors showed that the linear classifier suffers from non-uniform measurement errors. Especially, the error magnifies when the actual state is closer towards $\ket{1}$. This is partly due to the significant overlap zone between the $\ket{0}$ and $\ket{1}$ states created by the linear decision boundary. To circumvent this issue, the authors proposed two non-linear classifiers (based on circular and elliptical decision boundaries) and trained them to minimize the variance in measurement errors across different states. The authors reported a noticeable reduction in the variance of the errors over the linear classifier.

Another measurement calibration technique is available through IBM's Qiskit tool kit. In this method, bit-flip probabilities of various states are computed by running circuits with known outputs. Then, an output distribution is compensated with the knowledge of bit-flip probabilities to get a measurement error calibrated result. This method gives excellent results for circuits with a smaller number of qubits. However, for a larger number of qubits, it requires a prohibitively large number of circuits ($2^N$ circuits for N qubits) to generate the measurement calibration matrix. 

There are a few works on readout error mitigation using detector tomography~\cite{chen2019detector, Maciejewski2020mitigationofreadout}. 

\subsection{Leveraging extended native gates}
Most quantum computers offer only a single 2-qubit gate. For example, CNOT gate in IBM's superconducting qubits, Mølmer–Sørensen gate in IonQ's trapped-ion qubits, etc. However, new hardware is emerging which supports multiple 2-qubit gates~\cite{abrams2020implementation, filipp, bylander}. An extended native gate set can make the gate decomposition step more efficient. In~\cite{abrams2020implementation}, the authors note that a SWAP can be decomposed using two gates (1 CZ + 1 iSWAP) when both CZ and iSWAP gates are available as native instructions. If only CZ or only iSWAP is available as the native gate, it takes 3 CZ/iSWAP to decompose a SWAP. Therefore, an extended gate set can significantly reduce the gate count from SWAP insertions in NISQ architectures with limited connectivity. For a test case on QAOA circuits, they report $\approx30\%$ reduction in gate depth leveraging the extended gate set. In general, reducing gate count is desired to minimize decoherence and gate error for better resilience. 

\subsection{Leveraging program structure -- QAOA}
General-purpose quantum compilers apply generic rules to optimize any given quantum program. They do not take into account program-specific details for aggressive optimization. In \cite{alam-iccad-invited-qaoa, alam-micro, mahabubul-dac-qaoa}, the authors presented algorithm-specific compilation methodologies for QAOA which is a promising near-term algorithm.
In QAOA, the ZZ-interactions (can be implemented with 2 CNOTs and 1 RZ operation as in Fig.~\ref{fig:problem-encoding}) within a level are commutative \cite{alam-micro}. In other words, these operations can be re-ordered without affecting the output state of the circuit. 

 Parallelization of the ZZ-operations using binary bin-packing algorithm (Instruction Parallelization - IP), repeated compilation of QAOA-circuits with re-ordered layers guided by a branch-and-bound optimization heuristic (Iterative Compilation), layer-by-layer circuit construction and compilation prioritizing operations that require less SWAPs (Incremental Compilation - IC) or can be executed more reliably (Variation-aware Incremental Compilation - VIC) are proposed in \cite{alam-micro}. They also developed heuristics to manipulate QAOA-circuit properties to perform intelligent initial qubit allocation (Qubit Allocation and Initial Mapping - QAIM/ Variation-aware Qubit Placement - VQP). These suite of techniques achieved $\approx$53\% reduction in circuit depth, $\approx$23\% reduction in gate-count, and $\approx$63\% improvement in estimated success probability of QAOA-circuits over the existing state-of-the-art techniques. They also demonstrated $\approx$26\% improvement in performance on a real IBM device. Similar techniques can be developed for other near-term quantum algorithms to maximize performance and resource utilization.

\subsection{Error inclusive training of Quantum ML models}
The training of the quantum classifier leverages a quantum-classical hybrid loop where a PQC generates an output distribution, and a classical optimizer updates the parameters of the PQC based on the output to minimize a cost function (minimize loss during the training phase). The authors in~\cite{mahabubul-islped} noted that if the training is performed including noise in the PQC, the quantum classifier shows more resilient performance. The classification accuracy for a quantum parity classifier from~\cite{mahabubul-islped} is shown in Fig.~\ref{fig:parityIBM}. Here, ``pure'' is noiseless training, and ``noisy'' is noisy training with noise values of the respective day. The plot clearly shows training with noise has better accuracy compared to training without noise. However, noise values change over time. Therefore, the same trained classifier performs worse on a different day (``noisyDD''). As training a classifier every day is expensive, a reasonable trade-off is to use an average value for noise data collected over some time (``noisyAVG'').

\begin{figure}
 \begin{center}
    \includegraphics[width=0.48\textwidth]{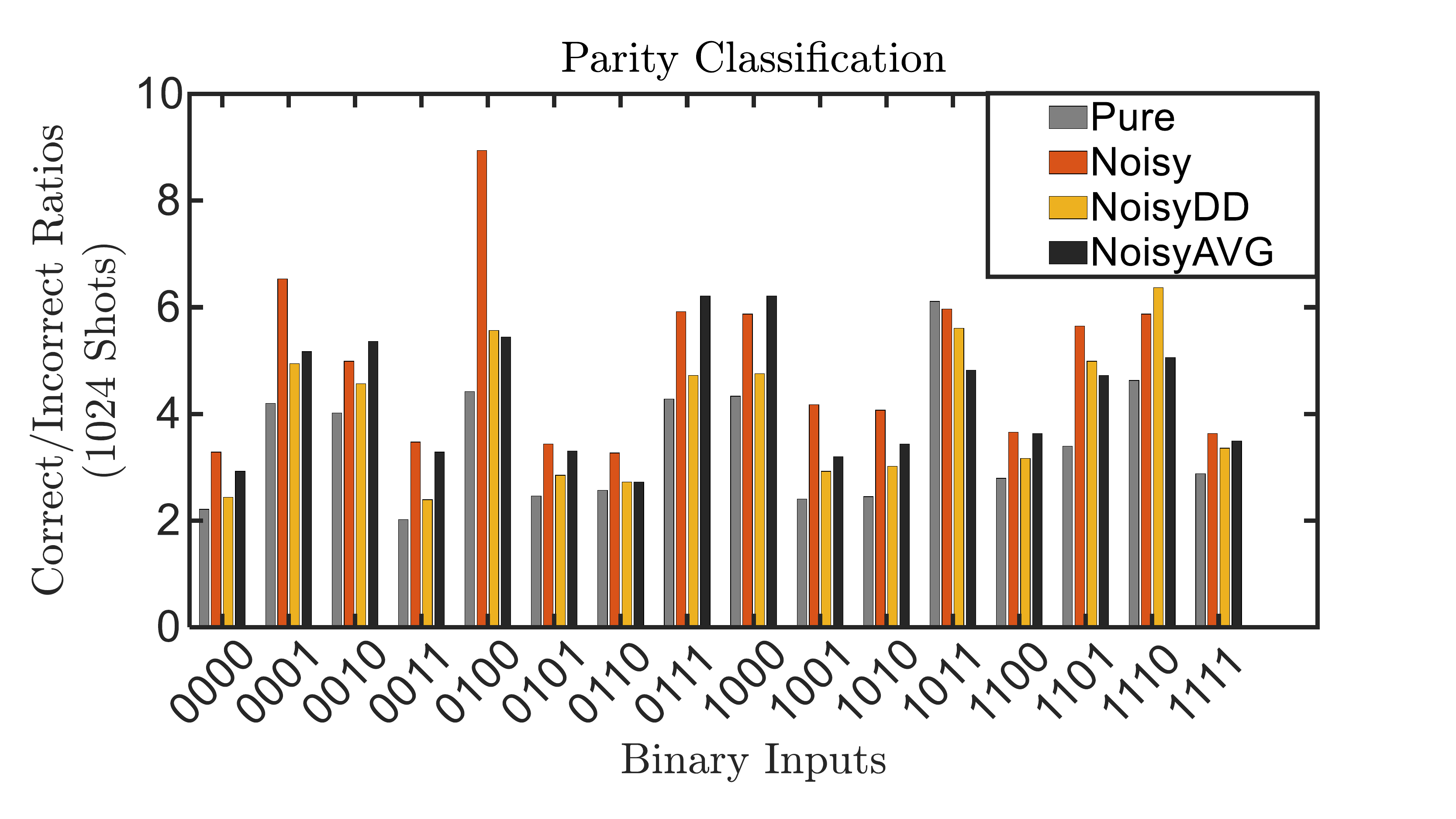}
 \end{center}
 \vspace{-1em}
 \caption{Inference performance of a 4-bit parity classifier (trained with three different approaches) on IBMQX4 \cite{mahabubul-islped}. \vspace{-5mm}} \label{fig:parityIBM}
 \label{fig:10} 
\end{figure}
\section{Countermeasures}\label{sec:countermeasures}
\subsection{Security attacks}
\subsubsection{Buffer qubits}
\begin{figure}
    \centering
    \includegraphics[width=3.2in]{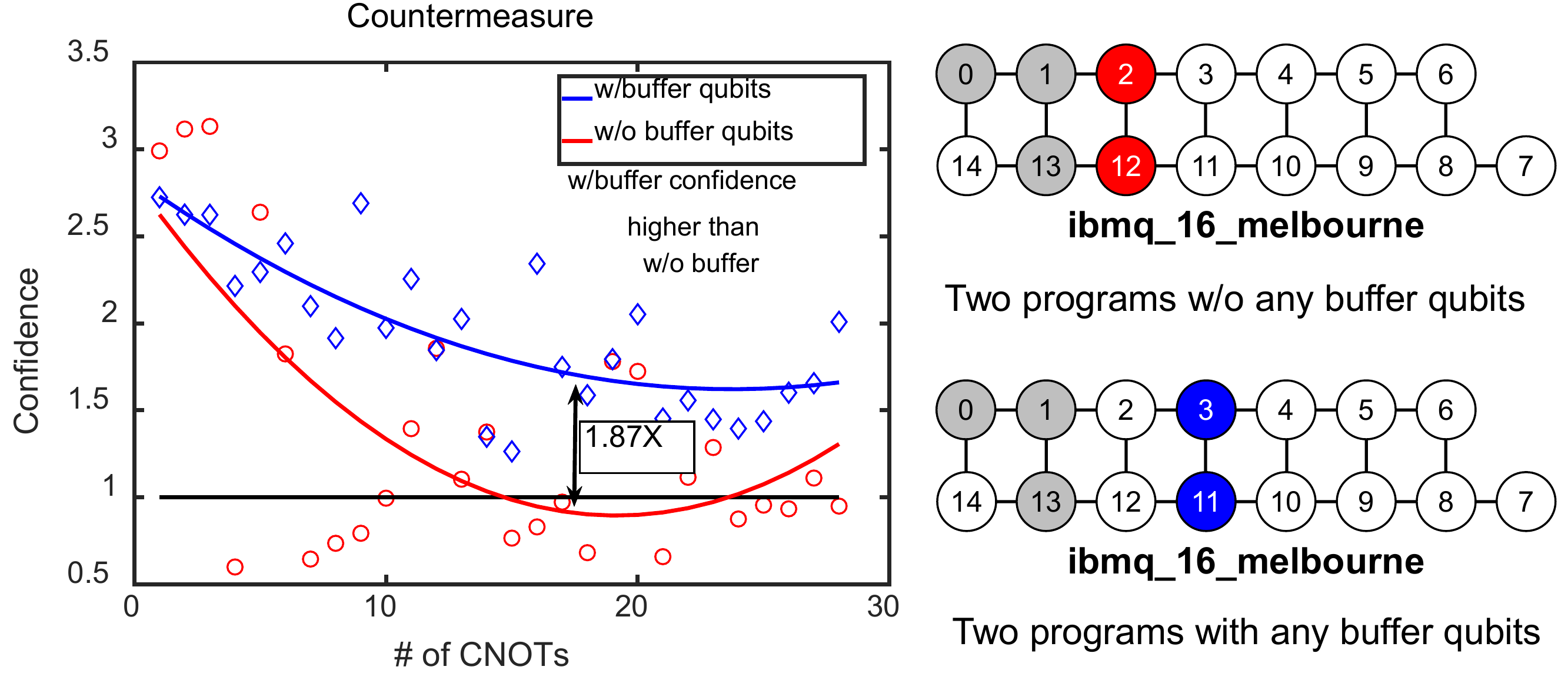}
    \caption{Buffer qubits as a countermeasure against crosstalk induced fault-injection attack.}
    \label{fig:buffer-qubits}
\end{figure}
The authors in~\cite{saki-islped} proposed using \emph{buffer qubits} to thwart crosstalk-induced fault injection. They experimentally demonstrated the countermeasure by running parallel circuits on \emph{ibmq\_16\_melbourne} another Canary processor (Fig.~\ref{fig:buffer-qubits}). In scenario--1, they allocated two programs on adjacent qubits: victim program on physical qubits \{Q0, Q1, Q14\} and adversary program on \{Q2, Q12\}. In scenario--2, they introduced buffer qubits between the programs where \{Q2, Q12\} acted as a buffer, and the adversary program was pushed to \{Q3,Q11\}. For scenario--2, they reported higher fidelity (as much as $1.87$x) than scenario--1.
\subsubsection{Authentication using QuPUF}
To verify whether the quantum hardware being allocated is the one that is desired or not, the authors in~\cite{puf} introduce the idea of Quantum PUF (QuPUF). A QuPUF is a quantum circuit that is sent to the quantum hardware. The parameters of QuPUF and the output given by the hardware act as the challenge-response pair respectively. For each hardware, the authors accumulate different challenge-response pairs. The assumption here is that each hardware will generate unique challenge-response pairs due to every hardware's unique characteristics like single-qubit error rates, CNOT error rates, decoherence time, and dephasing time.
They propose two models of QuPUF namely, Hadamard gate-based QuPUF and decoherence-based QuPUF (Fig. \ref{fig:qupufs}(a) and (b)) as described below.

\textbf{Hadamard gate-based QuPUF:}
The Hadamard gate-based QuPUF uses the biasing of the probability of the qubits towards either 0 state or 1 state to generate the response. The reason for such biasing could be  gate error (usually small for single-qubit gates) or readout error (typically large). At the start, all the qubit states are initialized to a zero state. They are then put in a superposition state using the Hadamard gate, and then the qubits are measured. Ideally, the output should be 50\% probability for both the states. But that won't be the actual case due to the errors and would be biased towards either 1 state or 0 state which would act as a unique device signature.

\textbf{Decoherence-based QuPUF:} 
The decoherence-based QuPUF relies on the decoherence times of the qubits to give unique output. The qubits are initialized to 0 state and then flipped to 1 state using a not gate. The qubits are then allowed to decohere down from 1 state to 0 state by the use of idle gates, which do no operation and simply pass time. In other words, the qubits are excited to a higher state and allowed to decohere down to 0 state. The decoherence of qubits will effectively act as the unique device signature.

\begin{figure}[h!]
\centering
    \begin{subfigure}[t]{0.20\textwidth}
        \centering
        \includegraphics[width=1.2in]{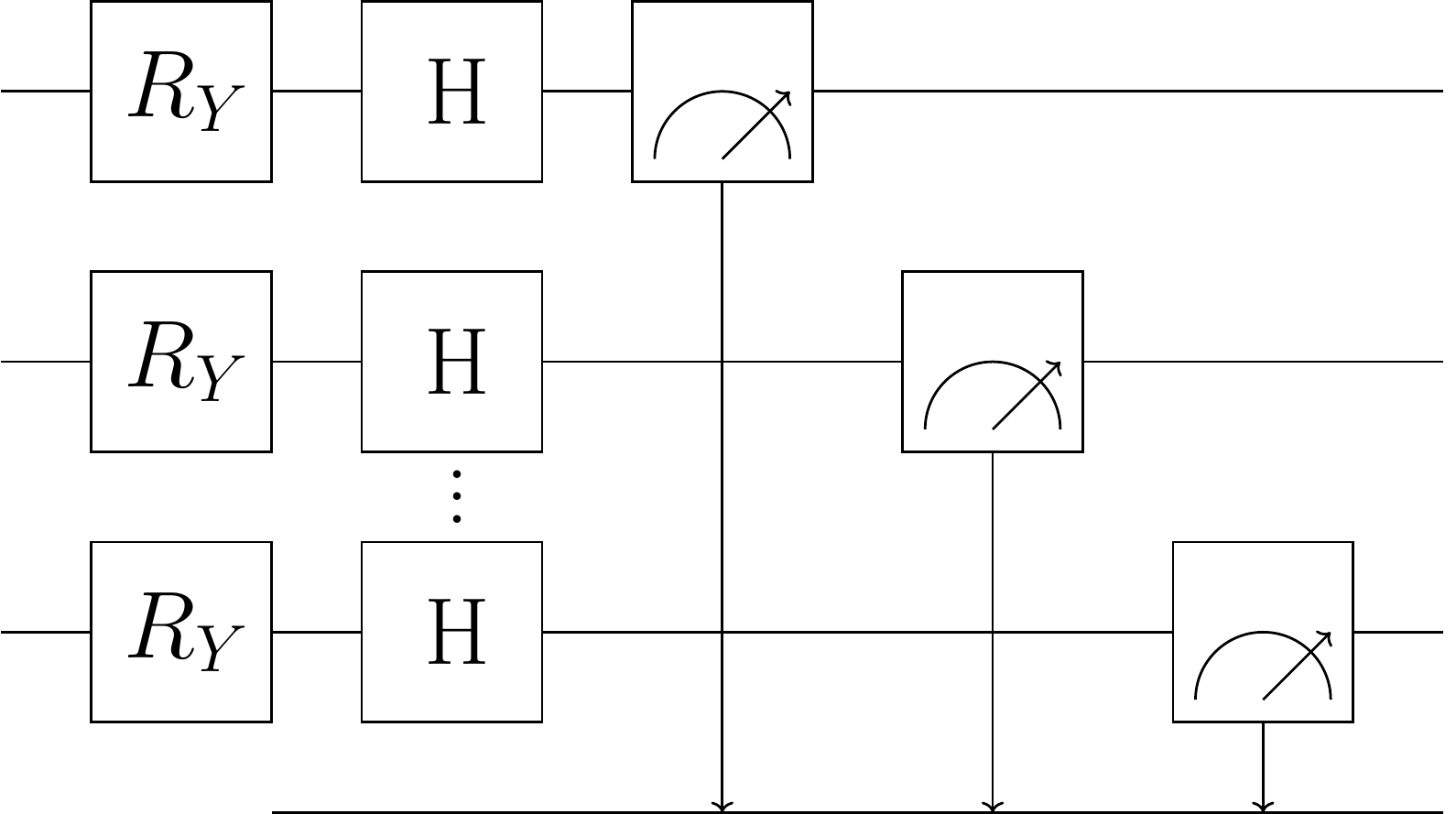}
        \caption{}
    \end{subfigure}%
    ~
    \begin{subfigure}[t]{0.30\textwidth}
        \centering
        \includegraphics[width=1.8in]{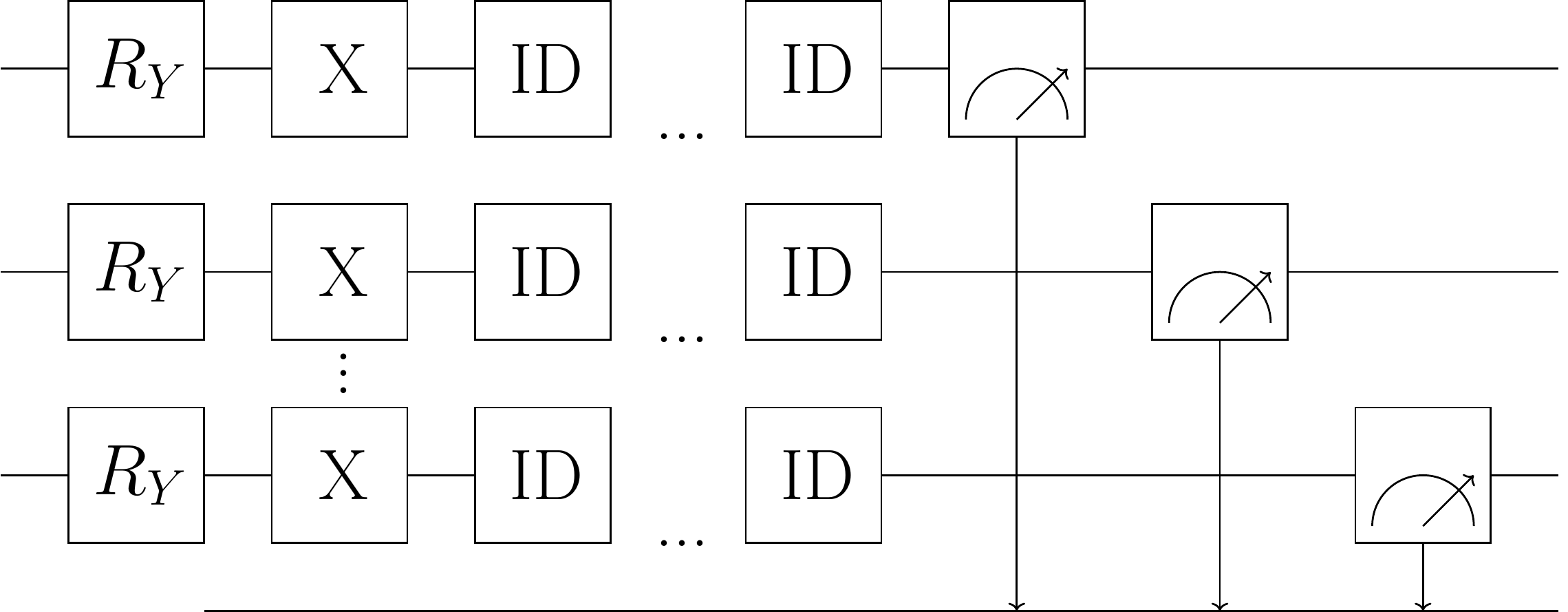}
        \caption{}
    \end{subfigure}
    
 \caption{Proposed QuPUFs: (a) Hadamard gate-based QuPUF; (b) decoherence-based QuPUF. The tunable rotation has been added for resilience.} \label{fig:f3}
 \vspace{-0.2em}
 \label{fig:qupufs}
\end{figure}
\begin{figure*} [t] 
 \begin{center}
    \includegraphics[width=0.9\textwidth]{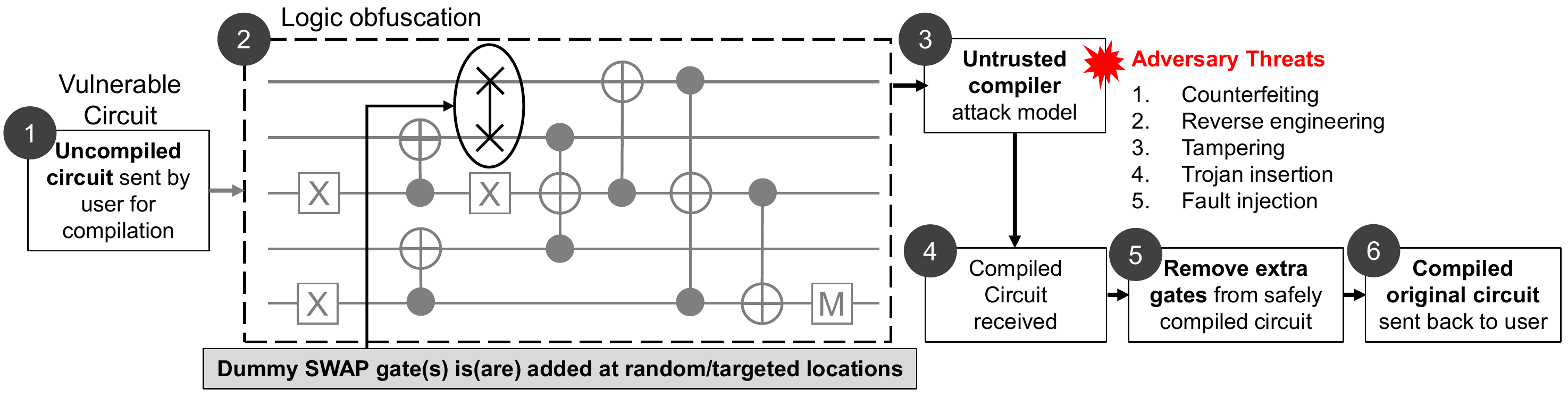}
 \end{center}
 \caption{Attack model proposed in~\cite{obfuscation-aks}. The quantum circuit is sent by the user to the untrusted compiler, where the adversary can steal the IP or reverse engineer the circuit. Logic obfuscation is proposed as countermeasure.} \label{fig:atkmodel}
\end{figure*}
\begin{figure*}
    \centering
    \includegraphics[width=0.80\linewidth]{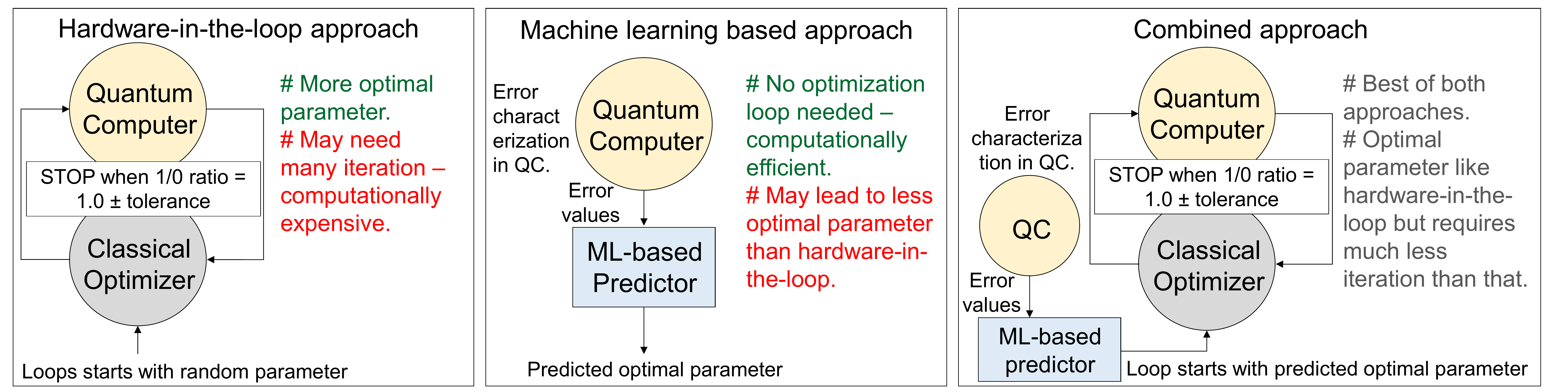}
    \caption{Overview and simplified flow of proposed methods for finding optimal gate parameter for the TRNG application.}
    \label{fig:trng}
\end{figure*}

\subsubsection{Tracking changes in error rates}
To detect unexpected changes in error-rates attack, the authors in~\cite{samah-iccad} proposed monitoring quantum circuit errors using \emph{test points}. They proposed three different types of tests: (i) classical test, (ii) superposition test, and (iii) un-compute test. A user needs to know the expected output to detect any changes in error rates. However, the user does not know output beforehand, otherwise the problem would become trivial. Besides, he/she cannot always resort to simulation since that is computationally expensive. Therefore, the tests are carefully chosen so that the user has knowledge about output. For example, the output of an un-compute test should be the initial state it started with. 
Two copies of a circuit with test points are run on two isomorphic sub-graphs of the device. The outputs are compared to check if the relative error rates are satisfied. If there is an anomaly, unexpected changes in compile-time information (error rates) are detected.


\subsection{Privacy attacks}
\subsubsection{Obfuscation using dummy gate}
As mentioned in Section~\ref{subsec:cloud}, the quantum circuit can be an IP. In~\cite{obfuscation-aks}, the authors proposed adding \emph{dummy gates} in a circuit to obfuscate the circuit from an untrusted compiler. The objective is to hide the true functionality of the circuit from the untrusted compiler. The adversary needs to identify and remove the dummy gates from an obfuscated circuit to extract the original circuit. This is a computationally hard problem since any gate can be a potential dummy gate. Any attempt to reuse the circuit without removing the dummy gates will result in corrupted or severely degraded performance.
Fig.~\ref{fig:atkmodel} conceptually shows the idea with a quantum circuit. The original circuit is divided into layers first. Then, inside each layer possible dummy SWAP insertion locations are identified. For example, if a layer has 3 free qubits, there are $\binom{3}{2} = 3$ choices for dummy SWAP gates. Therefore, there can be numerous SWAP insertion locations. However, only one dummy SWAP will be inserted in the original circuit and sent to the untrusted compiler.

The aim is to insert a dummy SWAP that will cause significant degradation in the output. The authors first ran exhaustive simulations with a set of test circuits and studied the impact of dummy SWAP insertion at each possible location. From the study, they developed a heuristic to find out an optimal SWAP insertion location. The heuristic tracks several features such as, the number of control qubits in the path from the SWAP to a measure qubit and calculates a score for the position. On the basis of the score, the optimal SWAP candidate is selected.

\subsubsection{Blind quantum computation}
Researchers considered protecting the privacy of quantum computation from compromised or malicious servers from an information-theoretic standpoint. The umbrella term for such line of research is \emph{blind quantum computation} (BQC)~\cite{arrighi2006blind}. Several theoretical protocols~\cite{bqc-1, bqc-2, mahadev2020classical} have emerged which allows a client to perform a computation on a server such that the server cannot learn any information about the client's input, output, and computation. Recently classical homomorphic encryption for quantum circuits has been proposed~\cite{mahadev2020classical}. The scheme allows a client to both hide data and performs computation on the hidden data. A review of the BQC protocols is presented in~\cite{bqc-review}. Although the theory for BQC is well researched, the physical implementations of such protocols are under-examined. A few works on physical implementations of BQC has been cited in~\cite{bqc-review}.

There is a recent work~\cite{samah-watermarking} on watermarking of quantum circuits. It includes a secret signature during the decomposition phase to watermark the IP~\cite{samah-watermarking}.

\subsection{Securing reversible circuits}
To circumvent the IC/IP piracy attack in~\cite{samah-ic-ip-piracy-tvlsi}, the authors presented two approaches. The first (naive) approach adds extra (dummy) ancillary and garbage lines \emph{pre-synthesis}. After the synthesis step, more ancillary and garbage lines will be added. The attack can identify only ancillary and garbage lines added post-synthesis, not the pre-synthesis ones. Thus, the first approach obfuscates the embedded functionality. However, adding extra lines pre-synthesis increases the hardware overhead.
To optimize the cost, they propose a second approach where reversible gates are added to the circuit in a judicious manner so that after synthesis the ``telltale'' signs are removed keeping the logical functionality intact.

To prevent piracy and reverse engineering from the end-use, the authors proposed logic locking~\cite{ets-limaye}. In particular, they chose SFLL-HD$^0$, a variant of \emph{stripped functionality logic locking} (SFLL) to secure the circuit. 
The logic locking block consists of 3 sub-blocks: functionally stripped circuit (FSC), restore unit/comparator, and restore signal/XOR. The FSC is formed by either adding or replacing a few logic gates. It inverts the output bit for one protected input pattern (PIP). The comparator/restore unit compares a key and the primary input to generate a restore signal. The key is saved in a tamper-proof memory. Finally, the XOR unit will revert the inverted output depending on the restore signal. The scheme protects against removal and SAT attacks.

\section{Security Opportunity: TRNG}\label{sec:opportunities}
Quantum superposition property can be used to generate true random numbers.
However, various noise sources, especially readout error, induce bias in the random bit generation. The authors in~\cite{saki-trng} proposed gate parameter optimization to compensate for this bias and to generate high quality true random numbers.

\subsubsection{Hardware-in-the-loop approach}
In this approach (Figure~\ref{fig:trng}), a real quantum computer starts with an initial gate parameter and generates a bitstream. A classical optimizer calculates the $1/0$ ratio from the stream and tries to minimize an objective function: $f = (1-ratio)^2$ by optimizing the gate parameter. The loop continues until an optimal parameter value is found which ensures a 1/0 ratio of $1.0~+ $ tolerance.  
Due to temporal variation, the optimal parameter varies with time and from device-to-device. Thus, the hardware-in-the-loop has to be invoked each time for each device which is tedious.

\subsubsection{Machine learning based approach}
In this approach, a statistical model (k-nearest neighbor regression) was trained to predict the gate parameter. First, the training data set is generated by running simulations in a noisy quantum circuit simulator. In the simulation, single qubit gate error, T1-relaxation time, and readout error are varied, and optimal gate parameter is computed for each combination. 
At the beginning, 3 qubit parameters - the gate error, T1-relaxation time, and readout error - are fed into the model, and the model outputs an optimal gate parameter. The true random number generator then starts with the parameter. 
This method avoids the slow classical optimizer-quantum hardware loop and quickly finds the optimal parameter. 

\subsubsection{Combined machine learning and hardware-in-the-loop approach}
In this approach, the previous two methods are combined to leverage the best of two approaches. First, the statistical model generates a \textit{near}-optimal parameter. Next, the hardware-in-the-loop starts from that parameter and finds the final (more) optimal parameter. As the optimization loop starts from a near-optimal point, the loop converges faster than random initialization.

\section{Future Outlook}\label{sec:outlook}

Quantum computers and the quantum workload possess several assets that could be lucrative targets for the adversaries to launch tampering, Denial-of-Service (DoS) attacks and/or to steal information. Quantum computers also come with several vulnerabilities including crosstalk and other noise sources, unencrypted netlist and sensitivity to noise, to name a few. Similar to classical computing, the security of quantum computing is intertwined with resilience. The presence of quantum computers in cloud and active usage are unlocking several new security and privacy threat vectors. The security of quantum computing requires an immediate attention to uncover various vulnerabilities and attack vectors so that proper defenses can be developed before the wide spread usage of this exotic and powerful technology.  

\section{Conclusions}\label{sec:conclusion}
In this paper, we review the challenges in NISQ devices. We discuss the state-of-the-art resilience techniques for addressing the challenges. We provide a survey of the security and privacy issues and their countermeasures in quantum computing. Finally, we discuss a projection into the future. As quantum computing matures and becomes ubiquitous, it will be more lucrative to attackers. Now is an opportune time to study the vulnerabilities and design safeguards for a resilient and secure future of quantum computing.

\bibliographystyle{IEEEtran}
\bibliography{IEEEabrv,ref}

\end{document}